\documentclass[conference]{IEEEtran} 
\usepackage[pdftex]{graphicx}
\usepackage{hyperref}
\usepackage{enumitem}
\usepackage{amsmath}
\usepackage{multirow}
\usepackage{multicol}
\usepackage{booktabs}
\usepackage[para,online,flushleft]{threeparttable}
\usepackage{amsfonts}
\usepackage{pifont}
\newcommand{\cmark}{\ding{51}}%
\newcommand{\xmark}{\ding{55}}%
\usepackage{xcolor, soul} 

\begin{document}

 \title{Detecting CAN Masquerade Attacks with Signal Clustering Similarity
\thanks{\scriptsize{This manuscript has been co-authored by UT-Battelle, LLC, under contract DE-AC05-00OR22725 with the US Department of Energy (DOE). The US government retains and the publisher, by accepting the article for publication, acknowledges that the US government retains a nonexclusive, paid-up, irrevocable, worldwide license to publish or reproduce the published form of this manuscript, or allow others to do so, for US government purposes. DOE will provide public access to these results of federally sponsored research in accordance with the DOE Public Access Plan (http://energy.gov/downloads/doe-public-access-plan).}}
}


\author{
\IEEEauthorblockN{Pablo Moriano\IEEEauthorrefmark{1}, Robert A. Bridges\IEEEauthorrefmark{2}, 
Michael D. Iannacone\IEEEauthorrefmark{2}}
\IEEEauthorblockA{\IEEEauthorrefmark{1}Computer Science and Mathematics Division, \IEEEauthorrefmark{2}Cyber Resilience and Intelligence Division \\ Oak Ridge National Laboratory\\
$\{$moriano, bridgesra, iannaconemd$\}$@ornl.gov}
}

\IEEEoverridecommandlockouts
\makeatletter\def\@IEEEpubidpullup{3.0\baselineskip}\makeatother
\IEEEpubid{\parbox{\columnwidth}{
    Workshop on Automotive and Autonomous Vehicle Security (AutoSec) 2022 \\
    24 April 2022, San Diego, CA, USA \\
    ISBN 1-891562-75-4 \\
    https://dx.doi.org/10.14722/autosec.2022.23028 \\
    www.ndss-symposium.org
}
\hspace{\columnsep}\makebox[\columnwidth]{}}



\date{}

\maketitle
\begin{abstract} \label{sec:abstract} 
Vehicular Controller Area Networks (CANs) are susceptible to cyber attacks of different levels of sophistication. Fabrication attacks are the easiest to administer---an adversary simply sends (extra) frames on a CAN---but also the easiest to detect because they disrupt frame frequency. To overcome time-based detection methods, adversaries must administer masquerade attacks by sending frames in lieu of (and therefore at the expected time of) benign frames but with malicious payloads. Research efforts have proven that CAN attacks, and masquerade attacks in particular, can affect vehicle functionality. Examples include causing unintended acceleration, deactivation of vehicle's brakes, as well as steering the vehicle. 
We hypothesize that masquerade attacks modify the nuanced correlations of CAN signal time series and how they cluster together. 
Therefore, changes in cluster assignments should indicate anomalous behavior. 
We confirm this hypothesis by leveraging our previously developed capability for reverse engineering CAN signals (i.e., CAN-D [Controller Area Network Decoder]) and focus on advancing the state of the art for detecting masquerade attacks by analyzing time series extracted from raw CAN frames. Specifically, we demonstrate that masquerade attacks can be detected by computing time series clustering similarity using hierarchical clustering on the vehicle's CAN signals (time series) and comparing the clustering similarity across CAN captures with and without attacks. We test our approach in a previously collected CAN dataset with masquerade attacks (i.e., the ROAD dataset) and develop a forensic tool as a proof of concept to demonstrate the potential of the proposed approach for detecting CAN masquerade attacks.

\end{abstract}
\section{Introduction} \label{sec:introduction}


Modern vehicles are complex cyber-physical systems containing up to hundreds of electronic control units (ECUs). ECUs are embedded computers that communicate over a (few) Controller Area Networks (CANs) to help control vehicle functionality, including acceleration, braking, steering, and engine status, among others.
CANs are vulnerable to cyber exploitation, both by adversaries with direct physical access (e.g., through the standard on-board diagnostic [OBD] II port) and remote access (e.g., Bluetooth, 5G). 
This increasing connectivity enables more advanced vehicle features at the expense of expanding the attack surface. 
By hijacking ECUs, attackers may stealthily manipulate CAN frames resulting in life threatening incidents. 
For example, malicious frame injection through cellular networks has resulted in unintended acceleration, vehicle brake deactivation, and rogue steering wheel turning~\cite{Miller:2015:Remote:Exploit:Black:Hat, Miller:2016:CAN:Exploit:Details}. 


CAN attacks are commonly classified using a three-tiered taxonomy that includes fabrication, suspension, and masquerade attacks~\cite{Cho:2016:Fingerprinting:ECUs, Verma:2020:ROAD}. Fabrication attacks inject extra frames, whereas suspension attacks remove benign frames; 
consequently, both categories usually disturb regular frame timing on the bus and can be accurately detected using time-based methods~\cite{Song:2016:IDS:Time:Interval, Moore:2017:Modeling:Inter:Time:IDS, Blevins:2021:Time-Based:CAN:IDS:Benchmark}. 
Masquerade attacks require the adversary to send frames in lieu of (and therefore at the expected time of) benign frames but with malicious payloads. In masquerade attacks, adversaries first suspend frames of a specific ID and then inject spoofed frames that modify the content of the frames instead of their timing patterns. Hence, masquerade attacks are the stealthiest CAN attacks. 



Masquerade attacks 
can still be detected because they alter the regular relationships of a vehicle's subsystems. Using an example attack from the ROAD~\cite{Verma:2020:ROAD} dataset, an adversary that gains control of the ECU(s) that communicate the wheel speed signals (four nearly identical signals) can modify the frames to break the near perfect correlation, which will stop the vehicle (regardless of the driver's actions). By understating the regular relationships of the vehicle's CAN signals, this condition can be flagged as anomalous, even if the modified signals are not abnormal when considered individually. 



The widespread dependence of modern vehicles on CANs, combined with the security vulnerabilities has been meet with a push to develop intrusion detection systems (IDSs) for CAN. Generally, there are two types of IDSs methods: signature and machine learning (ML). Signature-based methods rely on a predefined set of rules for attack conditions. Behavior that matches the expected signature is regarded as an attack~\cite{Larson:2008:Specification:IDS, Bresch:2017:Design:Implementation:IDS, Olufowobi:2019:Saiducant}. 
However, given the heterogeneous nature of the CAN bus in terms of transmission rates and broadcasting, effective rules for detecting attacks are difficult to design, which contributes to high rates of false negatives~\cite{Wu:2019:CAN:IDS:Survey}. 
In contrast, ML-based methods profile benign behavior to identify anomalies or generalized attack patterns when the traffic does not behave as expected. 


In doing this, many ML-based methods leverage the CAN's frame payloads~\cite{Taylor:2016:Anomaly:LSTM, Hanselmann:2020:CANET}. Note that in passenger vehicles, signals (sensor values communicated in CAN frames) are encoded into the frame payloads via proprietary (nonpublic, original equipment manufacturer-specified) mappings. Some IDSs operate on the binary payload (raw bits) \cite{Taylor:2016:Anomaly:LSTM}, whereas others operate on the time series of signal values \cite{Hanselmann:2020:CANET}.
Processing the binary payload has a set of associated challenges. 
First, there is a semantic gap with respect to the signals encoded in the payload. This means that a single CAN frame's payload usually contains several signals encoded in different formats, including byte ordering, signedness, label and units, and scale and offset~\cite{Verma:2021:CAN-D}. 
Second, detecting subtle masquerade attacks requires analyzing the payload content because the correlation between certain signals may change when the frame content is modified during an attack; hence, analyzing translated signals is a promising avenue. 
Thus, considering the relationship between signals is important for achieving a more effective defense against advanced masquerade attacks. 


In this work, we propose a forensic framework to decide if recorded CAN traffic contains masquerade attacks. The proposed framework works at the signal-level and leverages time series clustering similarity to arrive at statistical conclusions. In doing so, we use available and readable signal-level CAN traffic in benign and attack conditions to test our framework. The results obtained from our evaluation demonstrate the capability of the proposed framework to detect masquerade attacks in previously recorded CAN traffic with high accuracy. 
Our contributions in this paper are summarized as follows:

\begin{itemize}[leftmargin=*]
\item We detail a CAN forensic framework based on time series clustering similarity for detecting masquerade attacks. The proposed framework is based on (1) clustering time series using agglomerative hierarchical clustering (AHC); (2) computing a clustering similarity; and (3) performing hypothesis testing using the clustering similarity distributions to decide between benign and attack conditions. 

\item We perform a sensitivity analysis of detection capabilities with respect to the type of AHC used. 
We report our results and offer possible explanations. 

\item We evaluate the proposed framework on masquerade attacks from the ROAD dataset~\cite{Verma:2020:ROAD}. Evaluation results show very high effectiveness of detecting attacks of different levels of sophistication. Our results indicate that the proposed forensic framework can be built upon to yield a viable real-time IDS, but using these results to craft a short-time-to-detection IDS is future work.
\end{itemize}

\section{Related Work} \label{sec:related work}

Our research is informed by past work leveraging time series signal correlations for context characterization of cyber-physical systems. Here, we provide an overview of related work in this area.

Ganesan et al.~\cite{Ganesan:2017:Exploiting:Correlations:Heterogeneous:Sensors} introduced the notion of using pairwise correlations of vehicular sensor readings (e.g., speed, acceleration, steering) to characterize behavioral context. They used it for cluster analysis to identify distinct driver behaviors and detect potential attacks. 
Li et al.~\cite{Li:2017:IDS:Sensor:Correlation:Integration} leveraged correlations from multiple sensors to train a regression model that estimates a targeted sensor value. They used the difference between the estimated and observed sensor values as an anomaly signature. Sharma et al.~\cite{Sharma:2018:Pearson:Correlation:IDS} proposed to compute Pearson correlation matrices of geolocation-related signals (e.g., latitude, longitude, elevation, speed, heading) to estimate the state of neighboring vehicles and detect location forging misbehavior based on correlation matrices' distance. Guo et al.~\cite{Guo:2019:IDS:Sensor:Consistency:Frequency} proposed Edge Computing Based Vehicle Anomaly Detection, which focuses on analyzing the time and frequency domains of sensor data to detect anomalies. In the first step, they flag abrupt changes in the correlations of sensor readings in the time domain as an indication of anomalies. 
For more accurate anomaly detection in the second step, they further analyze the sudden change in sensor readings by computing the change in power spectral density (PSD) of sensor data in the frequency domain. Under anomalous circumstances, the PSD is expected to be higher in the high-frequency band. 
He at al.~\cite{He:2020:Exploring:Sensor:Redundancy:IDS} explored using correlations between heterogeneous sensors to identify consistency among sensor data (e.g., acceleration, engine RPM, vehicle speed, GPS) and then utilize the data to detect anomalous sensor measurements. They accomplished this by embedding the relationship of multiple sensors into an autoencoder and pinpointing anomalies based on the magnitude of the reconstruction loss.
Leslie~{\cite{Leslie:2021:Unsupervised:CAN:IDS}} developed an unsupervised learning method to detect malicious traffic over J1939 data. This method converts categorical features to numerical features with a one-hot encoding scheme and uses an ensemble AHC algorithm that integrates multiple linkage options.

Compared with the studies mentioned above, the present paper is unique in that we model temporal and signal-wise dependencies between CAN signals using time series clustering~\cite{Javed:2020:Time:Series:Clustering:Benchmark}. 
Specifically, we use AHC to generate a hierarchical relationship between signals known as a dendrogram~\cite{kaufman:2009:Clustering:Book}. 
Using a hypothesis test, we show that masquerade attacks are detectable by the resultant distribution of clustering similarities. In addition, our method is tested on real CAN data containing hundreds of signals, as opposed to previous methods that used a dozen signals at most.


\section{Methods} \label{sec:methods}


Our focus is on processing a set of $N$ signals (i.e., time series, $\mathcal{S} = \{X^{1}, X^{2}, \ldots, X^{N}\}$) obtained from a CAN log captured during a vehicle's drive. 
The subsections below explain the mathematical details of each step of our method and the data source used to perform this research. 
The proposed framework applies AHC (see \S~\ref{subsec:hierarchical clustering}) to produce a dendrogram of clusters of $\mathcal{S}$. 
Given two captures, each producing its corresponding dendrogram, we compute a similarity between the dendrograms using the CluSim method (see \S~\ref{subsec:clustering similarity})~\cite{Gates:2019:Clustering:Similarity}. Finally, the pairwise similarities from each capture's dendrograms are used to create a hypothesis test to distinguish between a benign CAN capture and an attack CAN capture. 

\subsection{Hierarchical Clustering} \label{subsec:hierarchical clustering}
\begin{figure}[!htb]
\setlength\abovecaptionskip{-0.1\baselineskip}
\centering
\includegraphics[width=1.0\columnwidth]{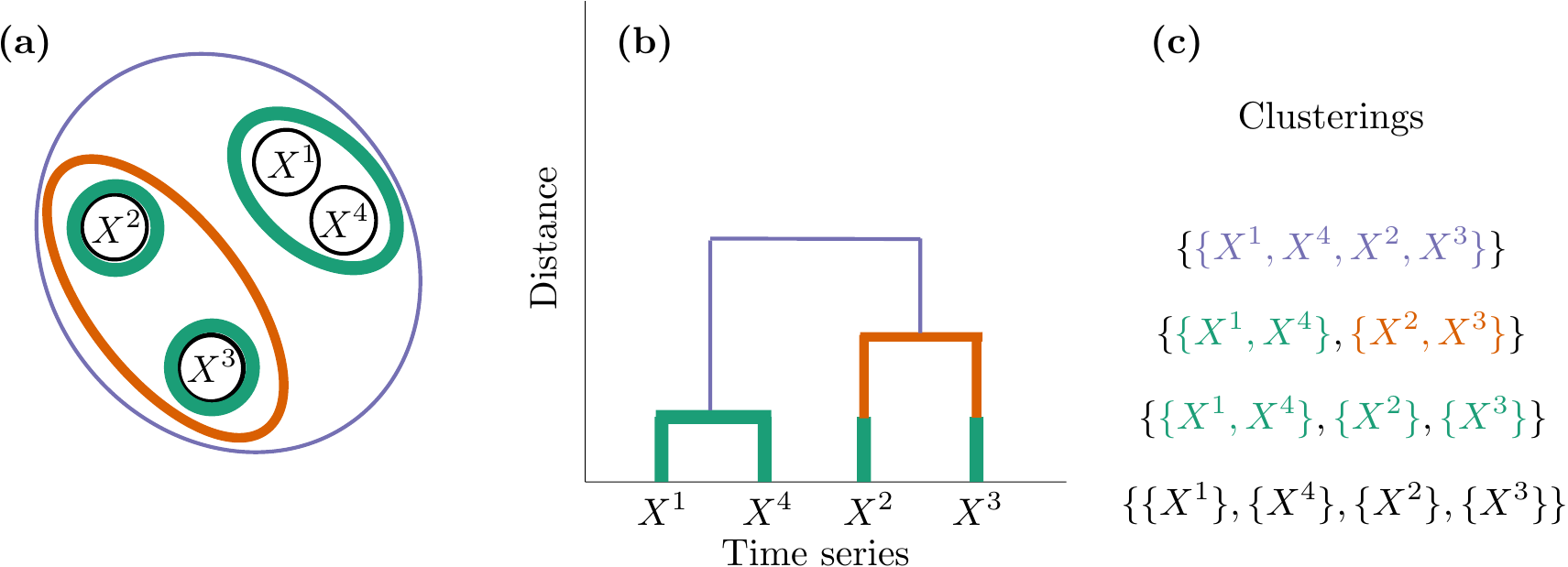} 
\caption{A hierarchical clustering using the agglomerative approach. 
(a) An example of a hierarchical clustering: here, $\{X^{1}, X^{2}, X^{3}, X^{4}\}$ is a set of time series to be clustered. 
They are grouped in a hierarchy of clusters by color (i.e., aquamarine, orange, purple). 
The thickness of the clusters represents how the hierarchy is built: close time series (aquamarine), more distant time series (orange), and most distant time series (purple). 
(b) Corresponding dendrogram of the hierarchical clustering depicted in (a). 
Time series are placed in the $x$-axis, and their relative distance is shown in the $y$-axis. Cluster colors correspond with the ones in (a). (c) Different clusters at each level of the hierarchy written explicitly. Note that cutting the dendrogram horizontally creates clusterings.}
\label{fig: AHC cartoon}
\end{figure}

Hierarchical clustering is a method that outputs a hierarchy of clusters (i.e., a set of nested clusters that are organized in a tree-like diagram known as dendrogram). 
It works by transforming a proximity matrix into a sequence of nested partitions. 
Figure~\ref{fig: AHC cartoon} depicts the details of a hierarchical clustering and its subsequent dendrogram using the agglomerative approach. The mathematical formulation of hierarchical clustering can be found in the Appendix~\ref{subsec: hierarchical clustering details}.
Agglomerative algorithms\footnote{There are two main categories of hierarchical clustering: agglomerative and divisive. Agglomerative places each object in its own cluster and gradually merges these atomic clusters into larger ones until all objects belong to a single cluster. Divisive reverses the process starting with all objects belonging to a single cluster and dividing them into smaller pieces~\cite{Lance:1967:Hierarchical:Clustering}. Here, we used the agglomerative approach by virtue of its simplicity.} 
require a definition of dissimilarity between clusters called a \textit{linkage}. The most popular linkages are the (a)~single linkage, which is the smallest dissimilarity between two points in opposite clusters; (b)~complete linkage, which is the largest dissimilarity between two points in opposite clusters; (c)~average linkage, which is the average dissimilarity over all points in opposite groups; and (d)~Ward's linkage, which focuses on how the sum of squares will increase when opposite groups are merged (or on the analysis of cluster variance). Ward's linkage tends to produce similar clusters as the $k$-means method~\cite{Javed:2020:Time:Series:Clustering:Benchmark}.  

Given a CAN capture that has been translated into its constituent signal time series,
$\mathcal{S} = \{X^{1}, X^{2}, \ldots, X^{N}\}$, we wish to cluster these time series to produce a dendrogram that represents their hierarchical structure. 
Each linkage choice (a)--(d) produces a potentially different dendrogram.
Understanding the most effective choice is a research question we address.

\subsection{Clustering Similarity} \label{subsec:clustering similarity}

\begin{figure}[!htb]
\setlength\abovecaptionskip{-0.1\baselineskip}
\centering
\includegraphics[width=1.0\columnwidth]{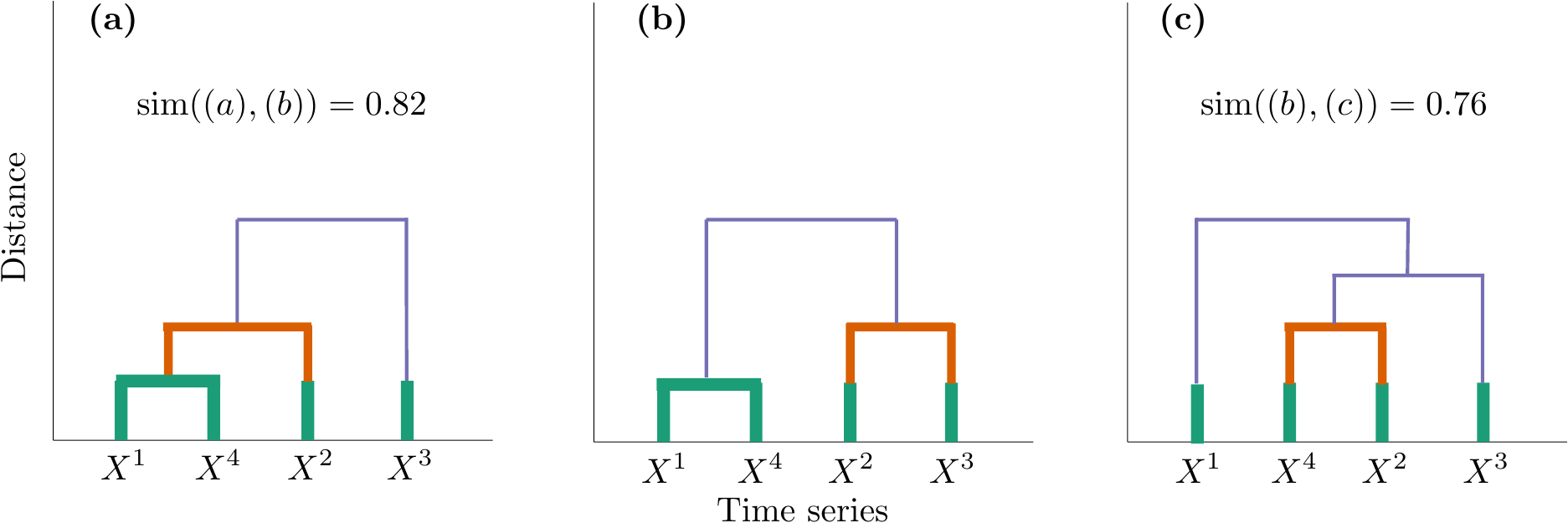} 
\caption{Hierarchical clustering similarity comparison using the CluSim method~\cite{Gates:2019:Clustering:Similarity}. 
Both (a) and (c) depict slightly modified dendrograms from that of (b). 
The similarity between (a) and (b) dendrograms is $0.82$, whereas the similarity between (b) and (c) dendrograms is $0.76$. 
This reflects that (a) and (b) are more similar than (b) and (c).}
\label{fig: similarity cartoon}
\end{figure}

Given two hierarchical clusterings (dendrograms) of a set $\mathcal{S}$, a clustering similarity quantifies a distance between them. We computed the similarity between dendrograms using the open-source CluSim method~\cite{Gates:2019:Clusim:Package}. The similarity value provided by this method exists in the range $[0, 1]$, where $0$ implies maximally dissimilar clusters, and $1$ corresponds to identical clusterings. 
We parametrized the clustering similarity method by letting $r=-5.0$ and $\alpha=0.9$. 
Figure~\ref{fig: similarity cartoon} shows a comparison between similarity scores of three dendrograms. 
The key advantage of CluSim is that it does not suffer from critical biases found in previous methods (e.g., normalized mutual information) and works for hierarchical clusterings, including in conditions of skew cluster sizes and a different number of clusters. We detail the main steps of CluSim in the Appendix~\ref{subsec: clusim details}. 

\subsection{Dataset} \label{subsec:dataset}

We used the ROAD dataset~\cite{Verma:2020:ROAD} to test our proposed forensic framework.
The ROAD dataset is an open set of CAN data collected from a real vehicle with fabrication attacks and a few advanced attacks (e.g., masquerade attacks).  
All of these attacks are physically verified (i.e., the effect of the CAN manipulation is observed and documented). 
Notably, masquerade attacks are also included but are simulated from the targeted ID fabrication attacks by removing the benign frames of the target ID. The ROAD dataset provides translated CAN time series following a similar schema used by Hanselmann~et~al.~\cite{Hanselmann:2020:CANET}. The fundamental advantage of using the ROAD dataset over prior released datasets is that it contains realistic, verified, and labeled attacks as opposed to synthetic ones. This opens the possibility for the evaluation, comparison, and validation of CAN signal-based IDS methods in realistic conditions.

We tested our forensic framework on the subset of masquerade attacks within the ROAD dataset. 
Each masquerade attack file in the ROAD dataset contains time series from hundreds of IDs that have a few to dozens of signals each. 
Table~\ref{Table: data description} shows the files we used from the ROAD dataset. 
Specifically, we tested the following attacks (in increasing order of detection difficulty): correlated signal, max speedometer, max engine coolant temperature, reverse light on, and reverse light off. 
In the correlated signal attack, the correlation of the four wheel speed values is altered by manipulating their individual values. 
In the max speedometer and max engine coolant attacks, the speedometer and coolant temperature values are modified to their maximum. 
In the reverse light attacks, the state of the reverse lights is altered to not match what gear the car is using (i.e., the reverse light is on when the vehicle is not in reverse, and the reverse light is off when the vehicle is in reverse). 

We used the complete set of 12 files to characterize the behavior in benign conditions ($\approx$3 hours of data). We used each of the files in the masquerade attack category: 3 files in the correlated attack ($\approx$1.3 minutes), 3 files for the max speedometer attack ($\approx$3.9 minutes), 1 file for the max engine coolant attack ($\approx$0.4 minutes), 3 files for the reverse light on attack ($\approx$3.2 minutes), 3 files for the reverse light off attack ($\approx$2.1 minutes) to characterize the attack conditions. 
\begin{table}[t]
\vspace{-1.7em}
\centering
\caption{CAN captures used from the ROAD dataset~\cite{Verma:2020:ROAD}.}
\label{Table: data description}
\scriptsize
\tabcolsep=0.1cm
\renewcommand{\arraystretch}{1.1}
\begin{tabular}{| c | l c c r |}
\hline
& \textbf{Description} & \textbf{\# Files} & \textbf{Used} & \textbf{Duration (min)} \\
\hline
\parbox[t]{2mm}{\multirow{3}{*}{\rotatebox[origin=c]{90}{Training}}} & Dynamometer Various Ambient & 10 & \cmark & 108.2 \\
& Road Various Ambient & 2 & \cmark & 70.6 \\
\cline{2-5}
& \emph{\textbf{Total}} & 12 & 12 & 178.8 \\
\hline
\parbox[t]{2mm}{\multirow{14}{*}{\rotatebox[origin=c]{90}{Testing}}} & Correlated Signal Fabrication Attack & 3 & \xmark & 1.3 \\
& Correlated Signal Masquerade Attack & 3 & \cmark & 1.3 \\
& Fuzzing Fabrication Attack & 3 & \xmark & 0.7 \\
& Max Engine Coolant Temp Fabrication Attack & 1 & \xmark & 0.4 \\
& Max Engine Coolant Temp Masquerade Attack & 1 & \cmark  & 0.4 \\
& Max Speedometer Fabrication Attack & 3 & \xmark & 3.9 \\
& Max Speedometer Masquerade Attack & 3 & \cmark & 3.9 \\
& Reverse Light Off Fabrication Attack & 3 & \xmark & 2.1 \\
& Reverse Light Off Masquerade Attack & 3 & \cmark & 2.1 \\
& Reverse Light On Fabrication Attack & 3 & \xmark & 3.2 \\
& Reverse Light On Masquerade Attack & 3 & \cmark & 3.2 \\
& Accelerator Attack (In Drive) & 2 & \xmark & 2.7 \\
& Accelerator Attack (In Reverse) & 2 & \xmark & 3.2 \\
\cline{2-5}
& \emph{\textbf{Total}} & 33 & 13 & 10.9 \\
\hline
\end{tabular}
\vspace{-1.5em}
\end{table}

\subsection{Pipeline Detailed Steps} \label{subsec:pipeline detailed steps}

The following steps are performed to arrive at the results. 

\subsubsection{Same Length Time Series Transformation} \label{subsubsec:time series transformation}

Each ID has a characteristic frequency that is unique in most cases. We modified the time series to have the same frequency by linearly interpolating them in common timestamps. We chose a baseline frequency of 10~Hz because it is the lowest frequency in the IDs in this dataset. This ensures that $\forall X^{i} \in \mathcal{S}, \ | X^{i}| = T$. Time series of the same length enable easier computation of similarity. 
After this, we also discarded any constant time series and normalized each remaining series to the unit norm.

\subsubsection{Time Series Correlation Computation} \label{subsubsec:time series correlation}

We computed pairwise Pearson correlations~\cite{Pearson:1895:Pearson:Correlation} among time series. 
Time series that have a positive correlation are expected to move in tandem (i.e., when one measurement increases or decreases, the other measurement also increases or decreases). 
Pearson correlation values that are close to $\pm 1.0$ indicate strong positive or negative correlation. 
As vehicle subsystems are dependent, we expect (1) clusters of correlated signals (e.g., increasing speed of the vehicle matches increases in the speedometer reading and the speed of all four wheels), and (2) such relationships to be broken or significantly changed upon a cyber attack.

\subsubsection{Hierarchical Clustering Computation} \label{subsubsec:hierarchical clustering}

These pairwise correlations populate a correlation matrix, which is used as the input for AHC.
The output is a dendrogram depicting hierarchies between clusters. 
We explored the effect of linkage selection (i.e., single, complete, average, Ward) in our detection framework.   

\subsubsection{Similarity Distribution Computation} \label{subsubsec:similarity distribution}

Once each dendrogram has been computed for each file, we compute empirical distributions of similarity between pairs of dendrograms using the method described in \S~\ref{subsec:clustering similarity}. 
We focus on two distinct groups. 
The first group is composed of all dendrograms derived from files in benign conditions (i.e., 12 files). 
In doing so, we computed pairwise similarities of dendrograms in this group, that is ${12 \choose 2} = 66$ possible combinations. 
The second group comes from the similarity between dendrograms in each category of attack (i.e., correlated, max speedometer, max engine coolant, reverse light on, reverse light off) and each of the files in benign conditions. 
This produces a varying number of combinations based on the number of files in each of the attack categories.

\subsubsection{Hypothesis Testing} \label{subsubsec:hypothesis testing}

We used the Mann-Whitney U test~\cite{Mann:1947:MannWhitneyUtest} and set the significance level to $0.05$ to test the null hypothesis that the distribution underlying benign conditions is the same as the distribution underlying attack conditions. 
The Mann-Whitney U test is a nonparametric test often used as a test of difference in location between distributions. 

\subsection{Motivational Preliminary Data Analysis} \label{sec:prelim} 

\begin{figure}[!htb]
    \centering
    \includegraphics[width=\linewidth]{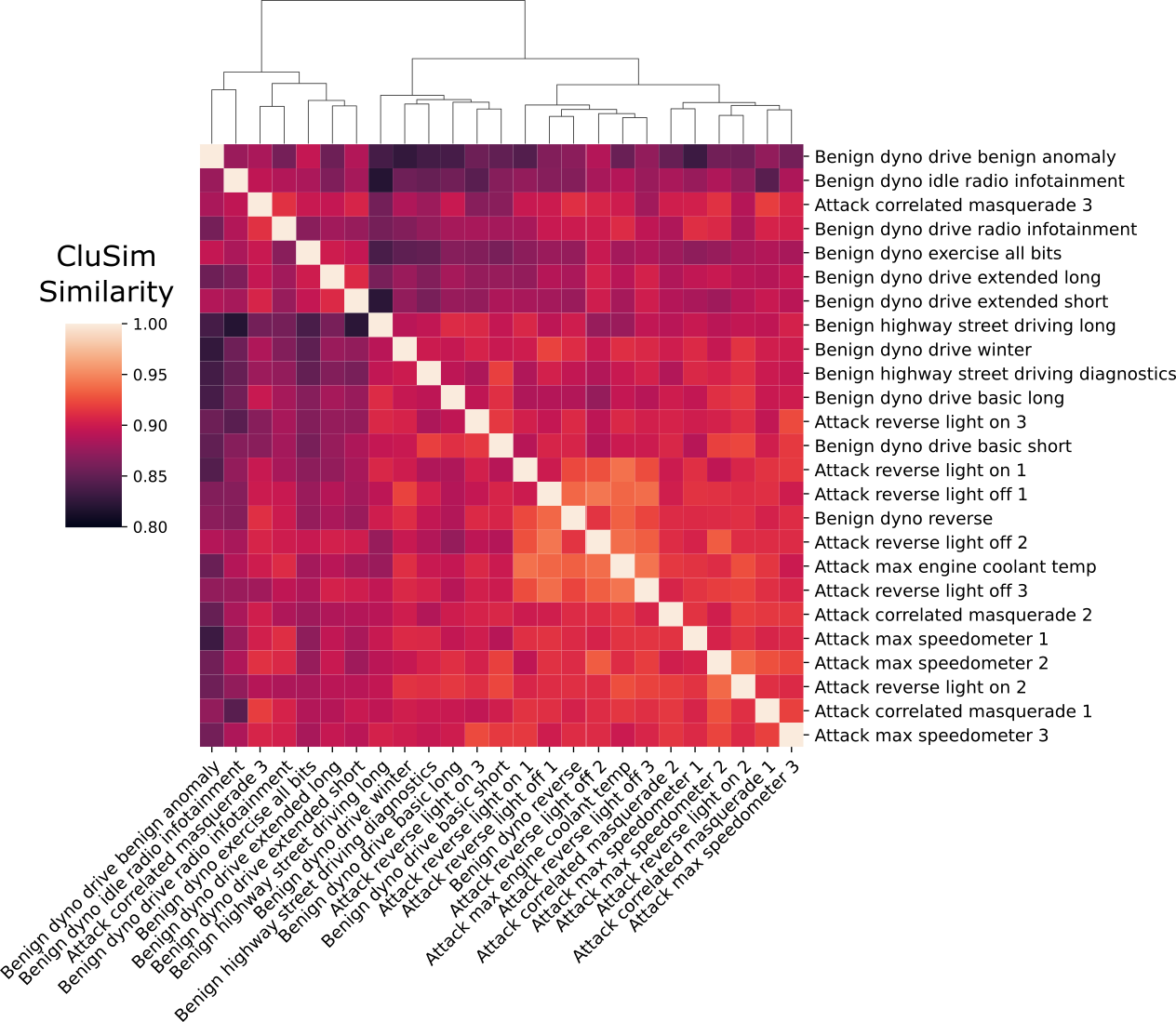}
    \caption{CluSim cluster similarity heatmap for each pair of files from the ROAD dataset (12 benign files, 13 masquerade attack files of five attack scenario types) depicted. For each file, a hierarchical clustering dendrogram is produced based on similarity of the file's CAN signals. For each pair of files, CluSim produces a similarity measure between the two file's dendrograms using Ward's linkage, $r=-5.0$, and $\alpha=0.9$, which is visualized by colors in the heatmap. 
    Atop the heatmap is a dendrogram showing hierarchical clustering of all files based on their the CluSim similarities. We used Euclidean distance on these similarities with Ward linkage. Notably, there are four main clusters. From left to right, the first and second contain all benign files except \texttt{Benign dyno reverse} file, and only two attack files, i.e., \texttt{Attack correlated masquerade 3} and \texttt{Attack reverse light on 3}, while the final two clusters contain all remaining attack files and the aforementioned benign file. As a preliminary analysis, this heatmap and clustering give positive results for masquerade attack detection by comparing Pearson correlation signal clusters.}
    \label{fig:clustermap}
    \vspace{-2.5em}
\end{figure}

As a first step to investigate our hypothesis that masquerade attacks will disrupt clustering based on correlation of the CAN signals, we compute and visualize the CluSim similarity (\S~ \ref{subsec:clustering similarity}) between every pair of files in the dataset discussed above (12 benign files and 13 masquerade attack files of five different attack scenarios, all in their signal time series format). 
More specifically, we follow the steps describe in \S~\ref{subsubsec:time series transformation} to interpolate time series to identical time steps, \S~\ref{subsubsec:time series correlation} to compute Pearson correlation for each pair of signals in a CAN file, and \S~\ref{subsubsec:hierarchical clustering} to produce a dendrogram for the signals in each file. 
We then apply CluSim, and visualize the pairwise similarity results in Figure~\ref{fig:clustermap}.

To see if the benign files signal cluster dendrograms do indeed ``look'' similar to each other but different than signal cluster dendrograms from masquerade attacks, we apply AHC to all the files based on their CluSim  similarities to each other. Figure~\ref{fig:clustermap} shows the resulting dendrogram revealing four main clusters. 
Notably the first two (from left to right) contain all but one benign file (11/12), and only two (of 13) attack files. 
This provides strong empirical motivation to pursue a more formal detection experiment based on hierarchical clustering of signals.

\section{Results} \label{sec:results}

Here we present our results on the efficacy of the proposed forensic framework for detecting masquerade attacks in the CAN bus. 
We focus on analyzing the detection capabilities for each of the attacks described in \S~\ref{subsec:dataset}. 
Figure~\ref{fig: correlated distributions} plots the probability density functions in the correlated attack (in benign and attack conditions) using the Gaussian kernel density estimate implementation from \texttt{seaborn}~\cite{Waskom:2021:seaborn} with a default bandwidth. We study the effect of the linkage selection (in the hierarchical clustering) for distinguishing between benign and attack conditions: (a)~single, (b)~complete, (c)~average, and (d)~Ward. 
We also report the $p$-value, using three decimals, of the associated Mann-Whitney U test to compare the two distributions in the inset; statistically significant values (i.e., $p\text{-value} < 0.05$) are printed in bold. 
Recall that we fixed the scaling parameter $r=-5$ for comparing hierarchical clusterings. 
This is because we want to capture differences at higher levels of the dendrograms, in which the focus is on coarser groups of multiple correlated signals, instead of more fine-grained groupings of individual to a few signals, in which not much emphasis is on their correlations.

Overall, we find that detecting attacks depends heavily on (1) the linkage function used to compute the hierarchical clusterings and (2) the severity of the attack in terms of the number of correlations perturbed. 
Specifically, out of the five attacks studied, the method based on Ward's linkage detected all of them (5~of~5), followed by complete linkage (4~of~5). Both single and average linkage methods detected fewer attacks (3~of~5). We report the $p$-values resulting from running the forensic framework for the remaining attacks (i.e., max speedometer, max engine coolant, reverse light on, and reverse light off) for each linkage in Table~\ref{table:p-values condensed}. We elaborate on each attack scenario below.
\begin{table}[t]
\centering 
\caption{Statistical Hypothesis Test Results ($p$-values).}
\label{table:p-values condensed}
\begin{threeparttable}
\begin{tabular}{l c  c  c  c}
\toprule
\textbf{Attack Scenario} & \textbf{Single} & \textbf{ Complete} & \textbf{Average} & \textbf{Ward}\\
\midrule 
Correlated & \textbf{0.005} & \textbf{0.002} & 0.123 & \textbf{0.000} \\
Max Speedometer & \textbf{0.003} & \textbf{0.017} & \textbf{0.007} & \textbf{0.000} \\
Max Engine Coolant & 0.251 & 0.065 & \textbf{0.006} & \textbf{0.008} \\
Reverse Light On & 0.378 & \textbf{0.007} & 0.057 & \textbf{0.004} \\
Reverse Light Off & \textbf{0.039} & \textbf{0.004} & \textbf{0.004} & \textbf{0.000} \\
\bottomrule
\end{tabular}
\begin{tablenotes}
    \item Statistically significant values are printed in bold.  
\end{tablenotes}
\end{threeparttable}
\end{table}

\subsection{Correlated Attack} \label{subsec:correlated attack}

\begin{figure}[!htb]
\vspace{-1.5em}
\setlength\abovecaptionskip{-0.5\baselineskip}
\centering
\includegraphics[width=0.8\columnwidth]{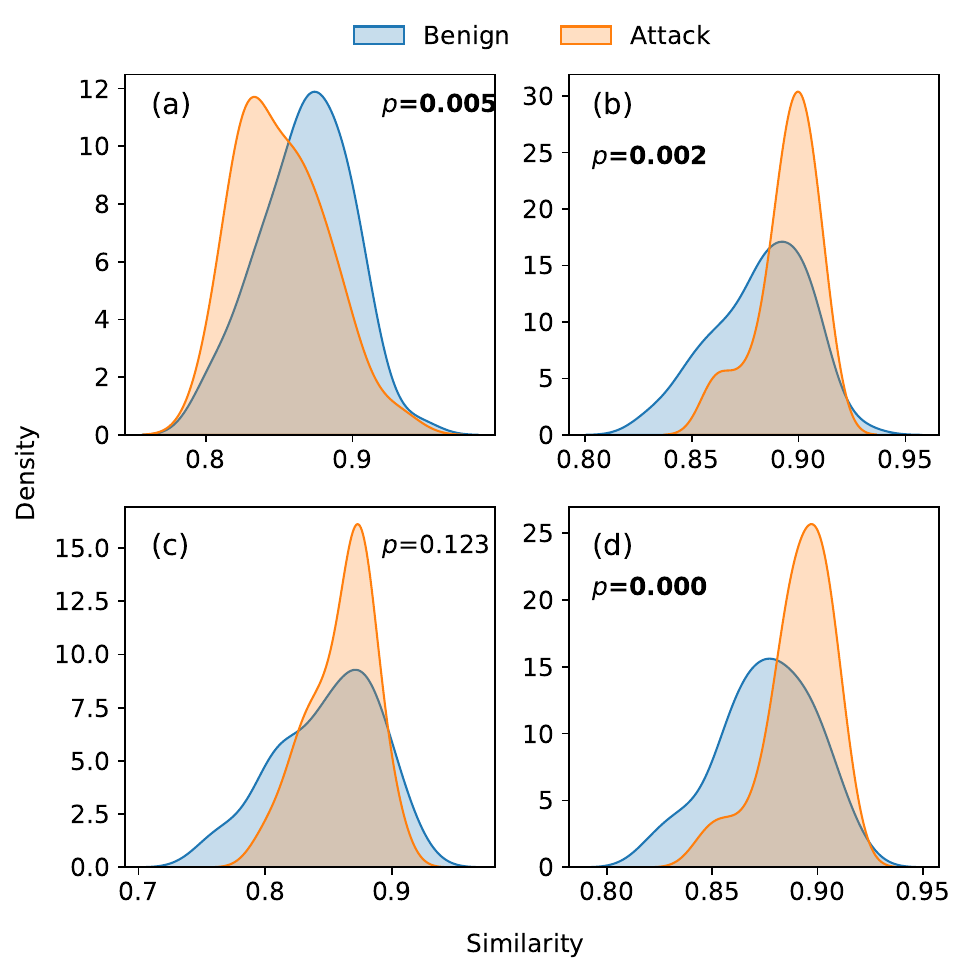}
\caption{Empirical distribution comparison of the correlated attack for each linkage selection: (a) single, (b) complete, (c) average, and (d) Ward. Results from these distributions appear in the first row of Table \ref{table:p-values condensed}.}
\label{fig: correlated distributions}
\end{figure}

Figure~\ref{fig: correlated distributions} shows the comparison of similarity distributions in the correlated attack. 
Among these, we found that the framework that used the average linkage (i.e., [c]) is not able to differentiate between benign and attack conditions. We also noticed that the Ward's method has the most distinctive difference (i.e., smaller $p$-value). 

\subsection{Max Speedometer Attack} \label{subsec:max speedometer attack}

Table~\ref{table:p-values condensed} shows that for the max speedometer attack, each linkage option produces statistically significant differences. We notice again that the Ward linkage produces the most distinctive results. We believe that speedometer readings correlate closely with wheel speed and engine readings, so when the speedometer value is manipulated (via attack) to appear maximally, correlations broken with these signals should be captured by the similarity distributions.

\subsection{Max Engine Coolant Attack} \label{subsec:max engine coolant attack}

Table~\ref{table:p-values condensed} shows the results of the max engine coolant temperature attack. We notice that only average and Ward linkages detect significant differences. In this attack, the engine coolant signal value is set to maximum, which may cause correlations with other engine signals to differ.

\subsection{Reverse Light On Attack} \label{subsec:reverse light on attack}

Table~\ref{table:p-values condensed} shows the comparison of similarity distributions in the reverse light on attack. Note that only the complete and Ward linkages produce statistically significant differences (i.e., [b] and [d]). We also note that, although statistically significant, these $p$-values are not as small as in the correlated attack, which is a consequence of having an attack that is more difficult to detect. This suggests that fewer correlated signals are affected under this attack, (i.e., only a binary [1 bit] signal was targeted).

\subsection{Reverse Light Off Attack} \label{subsec:reverse light off attack}

Table~\ref{table:p-values condensed} shows that for the reverse light off attack, each linkage method produces statistically significant differences. Among these, Ward's linkage produces the most significant difference, followed by average and complete linkages. The single linkage produces the least significant result, but it still meets the threshold.

\subsection{Detection Evaluation} \label{subsec:performance evaluation}

We compute and compare the performance of the proposed framework for classifying benign and attack files. Recall that we use 12 benign files. 
To do so, we implement a cross-validation as follows: 
We set apart three benign files to be used for testing purposes (along with all attack files) and use the remaining nine files for training, that is, for computing the similarity distribution from benign files. 
We chose to hold out three benign files for testing to be consistent with the maximum number of attack files found in the attack dataset (i.e., correlated, max speedometer, reverse light on, and reverse light off attacks each have three attack files). 
We implement the above train-test split of our benign files for each of the ${12 \choose 9} = 220$ possible combinations. 
This experimental design allows us to decide if the difference between similarity distributions in benign and attack scenarios is statistically significant and further count the number of true positives (TP), false positives (FP), false negatives (FN), and true negatives (TN). 

We use the best set of parameter values derived from our previous experiments, i.e., Ward linkage, $r=-5.0$, $\alpha=0.9$, and a significance level for the statistical hypothesis test of 0.05. 
We report the following micro-averaged classification metrics based on these numbers: 
\emph{Precision}, defined as $\frac{TP}{TP +FP}$, gives the likelihood that the computed similarity distribution difference can be attributed to an attack; 
\emph{Recall}, defined as $\frac{TP}{TP+FN}$, gives the likelihood that attack files are detected. 
Since higher precision often comes at the price of lower recall (and vice versa), it is important to consider a balance of both metrics, and the standard balanced metric is the F1 score, defined as $2 \times \frac{precision \times recall}{precision+recall}$. 
Table~\ref{table:sumary of results} summarizes these findings.
\begin{table}[t]
\centering 
\caption{Classification Results $(\%)$.}
\label{table:sumary of results}
\begin{threeparttable}
\begin{tabular}{l c  c  c }
\toprule
\textbf{Attack Scenario} & \textbf{Precision} & \textbf{Recall} & \textbf{F1 score}\\
\midrule 
Correlated & 88.00 & 100.00 & 93.62\\
Max Speedometer & 88.00 & 100.00 & 93.62\\
Max Engine Coolant & 87.18 & 92.73 & 89.87\\
Reverse Light On & 87.23 & 93.18 & 90.11\\
Reverse Light Off & 88.00 & 100.00 & 93.62\\
\bottomrule
\end{tabular}
\begin{tablenotes}
    \item False positive rate (FPR), defined as $\frac{FP}{FP+TN}$ equals to 13.64\%. Because this method is unsupervised, the training set is defined by the benign files in a given fold not the attack files; hence, the false positive counts and rate are independent of the attack scenarios. 
\end{tablenotes}
\end{threeparttable}
\end{table}

\begin{figure}[!htb]
    \vspace{-1.5em}
    \centering
    \includegraphics[width=1.0\columnwidth]{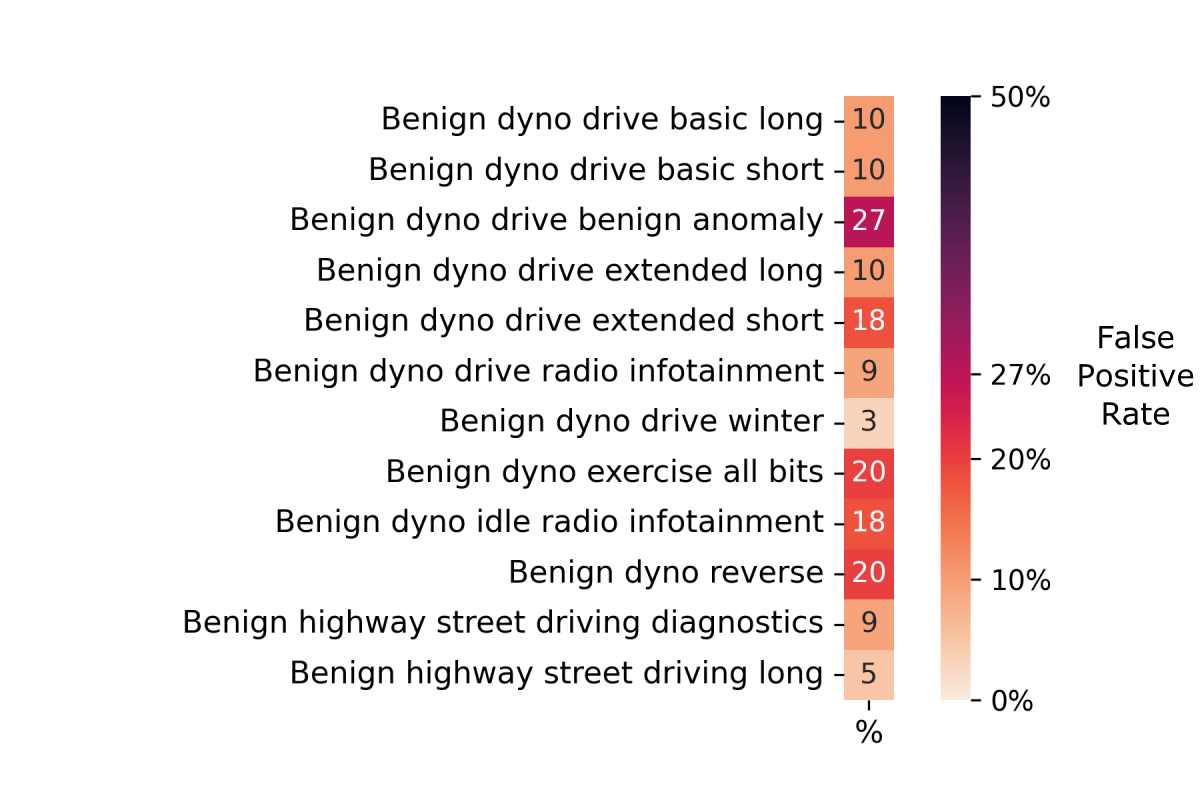}
    \caption{False positive rates for each benign file. All are at or below 20\%, except \texttt{Benign dyno drive benign anomaly} at 27\%. Comparing with Figure~\ref{fig:clustermap}, we see the lone benign file to cluster with attack files in preliminary analysis (i.e., \texttt{Benign dyno reverse}) has FPR = 20\%.}
    \label{fig:fpr}
\end{figure}

Because our method is unsupervised, the training set is defined by the benign files in a given fold not the attack files; hence, the false positive counts and rate are independent of the attack scenarios. 
In our experiment, the overall FPR is 13.64\%, with per-file FPR depicted in Figure~\ref{fig:fpr}.

Unlike the false positive and true negative counts, the true positive and false negative results will vary across different attack scenarios. 
For correlated, max speedometer, and reverse light off attacks, our results are identical. In these attack scenarios, recall is $100.00\%$ meaning that all the attack configurations are detected even when changing the set of benign files used in training. 
In these attack scenarios, precision is $88\%$ meaning that from the detected configurations, there are a few that come from the set of benign files or false positives. We obtain different results for the max engine coolant and reverse light on attacks. In particular, the lower results for the max engine coolant attack suggests that this attack was more difficult to detect when varying the set of benign files. 

\section{Discussion} \label{sec:discussion}

This work proposes a statistical forensic framework to detect masquerade attacks in the CAN bus. We quantify the empirical distribution of similarities of time series captures in benign and attack conditions. To accomplish this, we cluster time series using AHC and compute the similarity between their corresponding dendrograms. We find that masquerade attacks can be detected effectively using the proposed framework, and its discriminatory power depends on the linkage function being used in the AHC as well as the the impact of the attacks on correlated signals.

These results suggest that the proposed framework is a viable approach for detecting masquerade attacks in a forensic setting. We assume that the time series signal translation (or at least a high-fidelity translation) is readily available for use. This seems feasible with current and upcoming work in reverse engineering CAN bus signals, such as CAN-D~\cite{Verma:2021:CAN-D}. 

The proposed framework detects all masquerade attacks in the ROAD dataset when the Ward linkage is used. Note that Ward's linkage (d) is an appropriate choice in this context because it tends to produce dense-enough clusters and enables the capture of meaningful changes in clustering assignations when attacks occur.
In contrast, for the single linkage (a), clusters of signals tend to be spread out and often not compact enough with clusters having disparate elements. In the complete linkage (b), clusters of signals tend to be compact, but not far enough apart, with clusters having similar members. Additionally, for the average linkage (c), clusters tend to be relatively compact and relatively far apart, which strikes a balance between single and complete linkages. 

We note that the detection performance may also depend on specific attack features. Here, the detection difficulty is based on the potential number of correlated signals that are affected by the attack. Thus, an attack scenario in which wheel speed signals are modified, such as in the correlated attack, has a more noticeable effect of disrupting correlation with other signals than an attack that modifies the reverse lights because the wheel speed correlation attack manipulates four highly correlated signals (and seemingly strong correlations to many other signals), whereas the reverse light attacks modify a single signal that has correlation with gear selection but not many other signals.

Detection metrics are also affected by the number of files used to compute the similarity distribution. It other words, augmenting the number of files to estimate the similarity distribution helps to have better defined distributions that are later used for comparison purposes. This explains the lower results on the max engine coolant attack that contains a single file.

To the best of our knowledge, the results from this research are the first to show systemic evidence of a forensic framework successfully detecting masquerade attacks based on time series clustering using a dataset of realistic and verified masquerade attacks. The following are some limitations of our work.

\begin{itemize}[leftmargin=0cm]
\item[] \textbf{ROAD dataset conditions.} 
The ROAD dataset was collected on a single vehicle while being exercised mostly on a dynamometer. We acknowledge that more comprehensive data collection using different vehicles may be necessary to generalize our findings. 
We also are aware that driving conditions may affect correlations of CAN signals, and the dynamometer conditions may be restrictive.  

\item[] \textbf{Parameter tuning.} The proposed framework allows for flexible election of linkage functions (e.g., single, complete, average, Ward) for computing the hierarchical clusterings and the scaling parameter $r$ and $\alpha$ to control the influence of hierarchical clusterings with shared lineages. Here, we fixed the values of $r$ and $\alpha$ to focus on differences at higher levels of the dendrograms or in groups of correlated signals. However, we acknowledge that the optimal selection of these parameters may depend on the type of attack and driving conditions. We did not explore those variables in this research. 

\item[] \textbf{Not real-time detection.} As is currently presented, this is not a real-time detector.

\item[] \textbf{Baseline comparison.} We did not compare  our proposed forensic framework with other methods.

\end{itemize}


\section{Conclusion} \label{sec:conclusion}

In this research, we proposed a forensics framework for the detection of masquerade attacks in the CAN bus. 
To ascertain this fact in experiments, we compute time series clustering similarity. We show that the similarity of time series clusters under benign conditions exhibits statistically significant differences from the the similarity of time series clusters under attack conditions. 
We demonstrated these differences under different attack scenarios with different levels of sophistication using data from the ROAD dataset. 
This work shows that it is possible to detect masquerade attacks by effectively using the time series clustering representation of signals in the CAN bus and appropriate choices of parameters to group them.

Future work in this area includes the development of a real-time IDS that uses the principles described in this work. Additional work includes the translation of such developments to edge computing devices that can be integrated with real-world vehicle conditions.

\section{Acknowledgments} \label{sec:acknowlegments}

This research was sponsored in part by Oak Ridge National Laboratory's (ORNL's) Laboratory Directed Research and Development Program. This research used resources of the Compute and Data Environment for Science (CADES) at ORNL, which is supported by the Office of Science of the U.S. Department of Energy under Contract No. DE-AC05-00OR22725.

\small
\bibliographystyle{IEEEtran}
\bibliography{bibliography}

\appendix
\subsection{Hierarchical Clustering Definition} \label{subsec: hierarchical clustering details}

Here we mathematically define hierarchical clustering. A partition $\mathcal{P}$, of $\mathcal{S}$ breaks $\mathcal{S}$ into non-overlapping subsets $\{C^{1}, C^{2}, \dots, C^{m}\}$, i.e., $\mathcal{S} = \bigcup_{i \in \{1, 2, \ldots, m\}} C^{i}$. A clustering is a partition, so the elements of the partition are called clusters. A partition $\mathcal{B}$ of $\mathcal{S}$ is nested in a partition $\mathcal{A}$ of $\mathcal{S}$ if every subset of $\mathcal{B}$ is a subset of a subset of $\mathcal{A}$, i.e., $\forall C^{i} \in \mathcal{B} \ \exists j : C^{i}\subseteq C^{j} \in \mathcal{A}$. A hierarchical clustering is then a sequence of partitions in which each partition is nested into the next partition in the sequence.

\subsection{Brief CluSim Overview} \label{subsec: clusim details}

Here we describe how CluSim works in brevity. See Gates et al.~\cite{Gates:2019:Clustering:Similarity} for full details. Given $\mathcal{S} = \{X^{1}, X^{2} \dots, X^{N}\}$ and a clustering $\mathcal{A} = \{C^{1}, C^{2}, \dots, C^{m}\}$, first make the bipartite graph with elements of $\mathcal{S}$ on the left, clustering assignments from $\mathcal{A}$ on the right, and edges denoting containment (i.e., $(X^{i}, C^{j})$ is an edge if and only if $X^{i}$ is in cluster $C^{j}$). Note that this can be naturally extended to a dendrogram representing a hierarchical clustering $\mathcal{A}$ by using a weighted bipartite graph, where the weight of the edges is given by a hierarchy weighting function based on the level of the cluster assignation within the hierarchical clustering. 
Next, the bipartite graph is projected into the $\mathcal{S}$ elements producing a weighted, directed graph that captures the inter-element relationships induced by common cluster memberships. 
Now equipped with a weighted, directed graph on $\mathcal{S}$, the CluSim method captures high-order co-occurrences of elements by taking into account their paths to obtain an equilibrium distribution of a personalized diffusion process on the graph, or personalized PageRank (PPR)~\mbox{\cite{Haveliwala:2003:PPR}}, i.e., for each $X^i$ in $\mathcal{S}$, a PageRank version with restart to $X^i$ given by probability $1-\alpha$ is used to produced stationary distribution $\boldsymbol{p}_{i}$.
The element-wise similarity of an element $X^{i}$ in two different clusterings $\mathcal{A}$ and $\mathcal{B}$ is found by comparing the stationary distributions $\boldsymbol{p}^{\mathcal{A}}_{i}$ and $\boldsymbol{p}^{\mathcal{B}}_{i}$ using a variation of the $\ell^1$ metric for probability distributions. Finally, the similarity score of two clusterings $\mathcal{A}$, $\mathcal{B}$ is the average of element-wise similarities. CluSim is parametrized by specifying $r$ and $\alpha$. Here, $r$ is a scaling parameter that defines the relative importance of memberships at different levels of the hierarchy. That is, the larger $r$, the more emphasis on comparing lower levels of the dendrogram (zoom in). In addition, $\alpha$ is a parameter that controls the influence of hierarchical clusterings with shared lineages. That is, the larger $\alpha$, the further the process will explore from the focus data element, so more of the cluster structure is taken into account into the comparison. We used $r=5.0$ and $\alpha=0.9$ in Figure~\ref{fig: similarity cartoon}.

\end{document}